\documentclass[nolinenumbers]{aastex631}

\DeclareUnicodeCharacter{02BC}{<definition>}
\DeclareUnicodeCharacter{FF1A}{<definition>}

\begin{document}

\title{Attenuation of LHAASO PeVatrons by Interstellar Radiation Field and Cosmic Microwave Background Radiation}


\author{Jianli Zhang}
\affiliation{National Astronomical Observatories, Chinese Academy of Sciences, 100101 Beijing, Peopleʼs Republic of China}

\author{YiQing Guo}
\affiliation{Key Laboratory of Particle Astrophyics, Institute of High Energy Physics, Chinese Academy of Sciences, 100049 Beijing, Peopleʼs Republic of China\\
guoyq@ihep.ac.cn}
\affiliation{Tianfu Cosmic Ray Research Center, Chengdu 610213, Sichuan Province, China}

\begin{abstract}
"PeVatrons" refer to astrophysical sources capable of accelerating particles to energies around $10^{15}$ electron volts and higher, potentially contributing to the cosmic ray spectrum in the knee region. Recently, LHAASO has discovered a large number of PeVatrons, allowing us to investigate in greater depth the contributions of these sources to cosmic rays above the knee region. However, high-energy gamma rays undergo attenuation due to interactions with the interstellar radiation field and cosmic microwave background radiation, requiring corrections to restore the true spectral characteristics at the source. In this study, using interstellar radiation field model extracted from galprop code\citep{2022ApJS..262...30P}, we quantitatively calculated the spectral absorption effects of sources listed in the first LHAASO source catalog, with some sources showing absorption reaching 30\% at 100 TeV and 80\% at 3 PeV. We also calculated the high energy gamma ray absorption effects of Galactic microquasars, which are potential PeVatrons. By calculating the absorption effects, it will help differentiate the radiation mechanisms of the acceleration sources.
\end{abstract}


\section{Introduction} \label{sec:intro}
The origin of high energy cosmic rays(CRs) is a long standing unresolved issue in particle astrophysics. The energy spectrum of CRs can be described by a power-law spectrum of $E^{-2.7}$, extending up to the knee region at around 3 PeV\citep{2024PhRvL.132m1002C}, beyond which the spectrum softens\citep{1959bSov.Phys.JETP...35..441}. CRs below the knee region are believed to be produced and accelerated within the Milky Way\citep{1961PThPS..20....1G}, indicating the presence of PeV acceleration sources within our galaxy, known as PeVatrons\citep{2021Natur.594...33C,2024ApJS..271...25C}. Due to the influence of the interstellar magnetic field, CRs lose directional information about their sources, necessitating the study of CR origins through gamma rays produced by these CR sources. However, there are two competing mechanisms for generating gamma rays: one involves leptons producing gamma rays through inverse Compton scattering, while the other involves hadrons producing gamma rays through pion decay. Therefore, in order to distinguish between these two mechanisms, it is necessary to understand the characteristic energy spectrum of primary gamma rays.

In the very high energy(VHE) gamma ray band, i.e., above 100 GeV, extragalactic sources interact with extragalactic background light to produce electron-positron pairs, leading to the absorption and attenuation of gamma rays. In the 100 TeV energy range, the interstellar radiation field(ISRF) within our galaxy also causes significant absorption of gamma rays. The calculation of the absorption effects and their dependence on spatial distribution was initially performed by \citep{2006A&A...449..641Z} and \citep{2006ApJ...640L.155M} using the 2D Galactocentric symmetric ISRF model of \citep{2005ICRC....4...77P}. Later also other works investigated gamma ray galactic absorption due to pair production on ISRF and Cosmic Microwave Background Radiation(CMB) by computing the optical depth\citep{2016PhRvD..94f3009V,2017MNRAS.470.2539P,2018PhRvD..98d1302P}.Then \citep{2017ApJ...836..233G} study the absorption effect of gamma rays from Galactic center.

In the PeV energy range, gamma ray absorption is only related to the distance from the source to Earth, resulting in significant absorption effects for sources near the galactic center. In the field of VHE gamma astronomy, researchers have long focused on extragalactic sources in terms of cosmological distances. However, while theoretical studies have indicated the absorption effects of gamma ray sources within our galaxy, they have often been overlooked.

The primary reason the absorption effects of gamma rays in the Milky Way have been ignored for so long is the theoretical expectation that PeVatrons within our galaxy are rare. Many instruments have been unable to effectively observe 100 TeV sources, resulting in the longstanding inability to explore PeV astronomy. In recent years, this situation has changed significantly due to the large detection area and long duty cycle of the LHAASO (Large High Altitude Air Shower Observatory)\citep{2010ChPhC..34..249C}, which has identified 12 ultra high energy(UHE) gamma ray sources, with the highest energy reaching 1.4 PeV\citep{2021Natur.594...33C}. Recently, LHAASO released its first catalog, discovering dozens of PeVatrons throughout the Milky Way, including but not limited to PWNe, Binaries, and Microquasars\citep{2024ApJS..271...25C}. These findings from LHAASO provide a rich and crucial set of candidate samples for exploring the origins of PeV CRs within our galaxy.

In this article, we used the Galactic ISRF model extracted from galprop code\citep{2022ApJS..262...30P} to calculate the electron-positron pair absorption optical depth of LHAASO sources and calculated the absorption optical depth of potential high-energy source candidates in the Galaxy. We then applied the calculated optical depths to several typical PeV sources to obtain their intrinsic gamma ray energy spectra.

The paper is organized as follows: in Section 2, decribe the opacity of the Galactic ISRF and CMB, and our computation method. In Section 3, the gamma ray flux absorption of LHAASO PeVtrons are presented, and also showed the absorption of the potential PeVatrons -- the Galactic microquasars, then the ISRF corrected spectral of two sources. Finally, in Section 4, a concluding discussion and an outlook are given.

\section{Calculation method} \label{sec:style}
The ISRF is composed of the star light, the infra-red radiation. The VHE photons will suffer the attenuation due to the pair production $\gamma \gamma \rightarrow e^{-}e^{+}$, when travelling around the Galaxy. While the energis of the VHE gamma ray photons increasing to $\sim$ PeV, they interact with the homogeneous CMB. The relation of the original and observed flux of the source is
\begin{equation}\label{eq1}
F(E)=F_{0}(E)e^{-\tau{(E,L)}}
\end{equation}

where $F(E)$ represents the observed spectrum after the attenuation, $\tau$(E, L,b,l) represents the optical depth of the gamma rays, being a function of the gamma ray energy and the distance from the gamma ray source. $\tau(E, L, b,l)$ is composed of $\tau_{\gamma\gamma}^{ISRF}$ and $\tau_{\gamma\gamma}^{CMB}$. $\tau_{\gamma\gamma}^{ISRF}$ is given as：

\begin{equation}\label{eq2}
\tau_{\gamma\gamma}^{ISRF}{(E_{\gamma},L, b, l)}=\int dL \int dcos{\theta} \int dE_{ISRF}\frac{dn(E_{ISRF},L,b,l)}{dE_{ISRF}}  \sigma_{\gamma\gamma} \rightarrow e^{-}e^{+}(s)
\end{equation}

Where L is the line of sight parameter, so the integration of $dL$ is along the line of sight of the incoming gamma rays, which is related with the Galactic coordinate $(b,l)$. The $\frac{dn(E_{ISRF},L,b,l)}{dE_{ISRF}} $ is the number density of the ISRF photons at the cylindrical coordinates $(r, z)$, with the origin at the GC, which is extracted from the GALPROP code. We can got $r$ and $z$ by:
\begin{equation}\label{eq3}
r=\sqrt{R_{s}^2+L^2\cos^2 b-2R_{s}L\cos b \cos l}   
\end{equation}
and
\begin{equation}\label{eq4}
z=L\sin b
\end{equation}
where $R_{s}$ is the galactocentric radius of the Sun.

$\sigma_{\gamma\gamma \rightarrow e^{-}e^{+}(s)}$ is the cross section of the pair production, given by:
\begin{equation}\label{eq5}
\sigma_{\gamma\gamma \rightarrow e^{-}e^{+}(s)}=\sigma_{T}\cdot \frac{3m_{e}^2}{2s}[-\frac{p}{E}(1+\frac{4m_{e}^2}{s})+(1+\frac{4m_{e}^2}{s}(1-\frac{2m_{e}^2}{s}))log\frac{(E+p)^2}{m_{e}^2}]
\end{equation}
where $\sigma_{T}=8\pi^2\alpha^2/3m_{e}^2$ is the Thomson cross section for photon elastic scattering on a rest
electron, $m_{e}$ is the mass of electron, $p=\sqrt{E^2-m_{e}^2}$  and $E=\sqrt{s}/2$ are the magnitude of momentum and the energy of the electron at the center of mass system. At the laboratory system, $s$ is given by $s=2E_{\gamma}E_{ISRF}(1-cos\theta)$ with $\theta$ being the angle between the momentum of  the ISRF photon and the incoming $\gamma$ ray. To simple the case, we have assumed the ISRF photon is an isotropic distribution when doing the integration over the $\theta$.

The optical depth due to pair production on the CMB photons can be calculated similarly with formula \ref{eq2}. The integral along the line of sight is simple, since the photon number density can be same in any position. The optical depth for photons of energy $E_{\gamma}$ coming from a source at distance $L$ can be calculated as:
\begin{equation}\label{eq6}
\tau_{\gamma\gamma}^{CMB}{(E_{\gamma},L)}=L \int dcos{\theta} \int  n(E_{CMB}) dE_{CMB}  \sigma_{\gamma\gamma} \rightarrow e^{-}e^{+}(s)
\end{equation}

\section{Calculation results} \label{sec:results}
Based on above method, we selected the PeVatrons in LHAASO first catalogue, and potential PeVatrons of microquasars in LHAASO field of view, then calculated the absorption effect of gamma ray emission from these sources. 

\subsection{Absorption of LHAASO PeVatrons} \label{subsec:results_LHAASO}
The VHE gamma rays will interact with ISRF and CMB photons, through pair production, the probability depend on the cross section. In general, If we know the distance of the sources from the Earth, with the ISRF and CMB models, and the cross section of the pair production, the optical depth of the gamma rays can be calculated. The expected attenuation down to a peak at around 3 PeV, due to the gamma rays interacting with CMB photons.

The first catalog of VHE and UHE gamma ray sources detected by the LHAASO has released\citep{2024ApJS..271...25C}. This catalog covers declination from $-20^{\circ}$ to $ 80^{\circ}$. It is the most sensitive large coverage gamma ray survey of the sky above 1 TeV. The catalog contains 90 point sources, with a significance of detection higher than $5\sigma$. The majority of these LHAASO sources are expected to be Galactic sources, due to their extended properties or their KM2A component detection. The LHAASO catalog contains a rich Galactic gamma ray emitters, such as SNRs, PSRs and their PWNe, massive star clusters, star-forming regions, superbubbles, binaries, etc.

Based on this catalog, we select the sources association with the known sources which has distance measured. We also exclude the extragalactic sources. After selection, there are 30 sources left, details see table \ref{tab:LHAASO}. The distance of some sources exceed to 8 kpc. 

\begin{deluxetable*}{lcccccccccccccc}
\tabletypesize{\scriptsize}
\tablewidth{0pt} 
\tablecaption{Selected LHAASO PeVatrons \label{tab:LHAASO}}
\tablehead{
\colhead{Source} & \colhead{Potential counterpart}& \colhead{Type} & \colhead{Dist(kpc)} &
\multicolumn{2}{c}{Coordinate} \\
\cline{5-6}
\colhead{} & \colhead{} &
\colhead{} & \colhead{} & \colhead{$l$} & \colhead{$b$}
} 
\startdata 
{1LHAASOJ0634+1741u}& Geminga& PWN & 0.25 &   195.25 & 4.3   \\ 
\hline
{1LHAASOJ1740+0948u}& PSR J1740+1000 & PSR & 1.23 &   33.790 & 20.26   \\
 \hline
{1LHAASOJ1839-0548u} & PSR J1838-0537 & PSR &1.3 & 26.36 & 0.07 \\ 
 \hline
{1LHAASOJ2031+4127u} & PSR J2032+4127 & PWN &1.33 & 80.18 & 1.09 \\ 
 \hline
{1LHAASOJ0007+7303u} & CTA 1  & PWN &1.4 & 119.71 & 10.47 \\ 
 \hline
{1LHAASOJ1825-1256u} & PSR J1826-1256 & PSR &1.55 & 18.51 & -0.29 \\ 
 \hline
{1LHAASOJ0542+2311u} & PSR J0543+2329 & PSR &1.56 & 184.57 & -3.53 \\ 
 \hline
{1LHAASOJ2020+3649u} & PSR J2021+3651 & PSR &1.8 & 75.18 & 0.12 \\ 
 \hline
{1LHAASOJ1959+2846u} & PSR J1958+2845 & PSR &1.95 & 65.94 & -0.43 \\ 
 \hline
{1LHAASOJ1954+2836u} & PSR J1954+2836 & PSR &1.96 & 65.23 & 0.40 \\ 
 \hline
{1LHAASOJ0534+2200u} & Crab & PWN &2.0 & 184.55 & -5.80 \\ 
 \hline
{1LHAASOJ1908+0615u} & PSR J1907+0602 & PSR &2.37 & 40.41 & -0.86 \\ 
 \hline
{1LHAASOJ2228+6100u} & SNR G106.3+02.7 & SNR &3.0 & 106.42 & 2.81 \\ 
 \hline
{1LHAASOJ0216+4237u} & PSR J0218+4232 & PSR &3.15 & 139.17 & -17.55 \\ 
 \hline
{1LHAASOJ1809-1918u} & PSR J1809-1917 & PSR &3.27 & 11.07 & 0.12 \\ 
 \hline
{1LHAASOJ1857+0203u} & SNR G035.6-00.4 & SNR &3.6 & 35.46 & -0.41 \\ 
 \hline
{1LHAASOJ1825-1337u} & PSR J1826-1334 & PSR &3.61 & 17.91 & -0.62 \\ 
 \hline
{1LHAASOJ1831-1007u} & PSR J1831-0952 & PWN &3.7 & 21.61 & -0.12 \\ 
 \hline
{1LHAASOJ1852+0050u} & PSR J1853+0056 & PSR &3.84 & 33.79 & 0.17 \\ 
 \hline
{1LHAASOJ1928+1746u} & PSR J1928+1746 & PSR &4.34 & 52.92 & 0.15 \\ 
 \hline
{1LHAASOJ1912+1014u} & PSR J1913+1011 & PSR &4.61 & 44.49 & -0.04 \\ 
 \hline
{1LHAASOJ1848-0153u} & W 43 & Massive Star Cluster &5.5 & 30.89 & -0.15 \\ 
 \hline
{1LHAASOJ1928+1813u} & SNR G053.4+00.0 & SNR &6.0 & 53.28 & 0.42 \\ 
 \hline
{1LHAASOJ2002+3244u} & SNR G069.7+01.0 & SNR &6.46 & 69.70 & 1.03 \\ 
 \hline
{1LHAASOJ1837-0654u} & SNR G24.7+0.6 & PWN &6.6 & 25.21 & -0.08 \\ 
 \hline
{1LHAASOJ1848-0001u} & IGR J18490-0000 & PWN &7.0 & 32.61 & 0.59 \\ 
 \hline
{1LHAASOJ1929+1846u} & SNR G054.1+00.3 & PWN &7.0 & 53.88 & 0.45 \\ 
 \hline
 {1LHAASOJ1959+1129u} & 4U 1957+11 & LXB &9.4 & 51.10 & -9.42 \\ 
 \hline
{1LHAASOJ1843-0335u} & SNR G28.6-0.1 & SNR &9.6 & 28.84 & 0.09 \\ 
 \hline
{1LHAASOJ1914+1150u} & PSR J1915+1150 & PSR &14.01 & 46.13 & 0.26 \\ 
\enddata
\tablecomments{We take the sources characteristics from \citep{2024ApJS..271...25C}. Noted that some sources not firmly identified, but association with pulsars, and the source may powered by the pulsar, so we used the distances of the association pulsars based on \citep{2024ApJS..271...25C}. }
\end{deluxetable*}

\begin{figure}[ht!]
\plotone{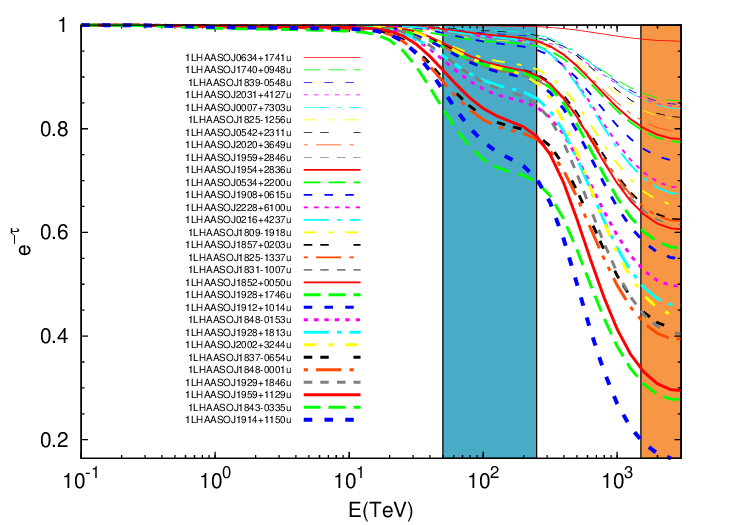}
\caption{Attenuation of the VHE gamma rays from LHAASO sources due to interaction with the ISRF and CMB as a function of the ray energy.  The labels in right axis represent VHE gamma rays from the distance of the sources interaction with CMB. The blue rectangle region is the ISRF interaction region. The yellow rectangle represent the CMB interaction effect.
\label{fig1}}
\end{figure}

Figure \ref{fig1} show the attenuation of gamma rays from these sources as a function of gamma ray energy.The cutoff begins at about 20 TeV, due to ISRF component. It is significant that the attenuation can reach up to 20\% for source 1LHAASO J1959+1129u with distance 9.4 kpc, mainly due to ISRF, while it can reach as much as 70\% at about 3 PeV, due to CMB component. The attenuation of gamma rays from the sources close to Galactic center is consistent with the results from the calculation before. These attenuation of the LHAASO PeV sources from larger distance is first clearly pointed out. The attenuation of VHE sources with distance larger than 8 kpc is a bigger effect, due to the ISRF, while for PeVatrons, the attenuation are huge effect. The study of these sources characteristics should be corrected by these huge effects.

\subsection{The absorption of Galactic microquasars\label{subsec:figures}}

LHAASO has observed dozens of sources of photons above 100 TeV. The possible sites of PeV radiation are supernova remnants, pulsar wind nebulae, young stellar clusters and superbubbles. LHAASO has renewed the knowledge that supernova remnants were believed to be the main sites where Galactic CRs originate. 
Microquasars, which are X-ray binaries with relativistic jets and stellar mass analogs of quasars, have been proposed as an additional candidates of the origins of high energy CRs( \citep{1984Natur.312...50H,2002A&A...390..751H,2004A&A...423..405B}). They are efficient particle accelerators to produce non-thermal emission up to PeV\citep{2022A&A...665A.145E}. Up to now\citep{2024arXiv240504469T}, only two microquasars have been detected in TeV domain:SS 433\citep{2024Sci...383..402H} and V4641 Sgr\citep{2023JPhCS2429a2017T}. Also, for Cygnus X-1, there is a strong hint of transient emission was discoveried\citep{2007ApJ...665L..51A}. In the Cygnus region, which is the highest energy photon -- 1.4 PeV photon located, Cygnus X-3 is also a potential candidate of origin of these PeV emission. So these microqusars will be very important targets to observation of LHAASO in future. It's important to study the attenuation effect of these microquasars.

\begin{figure}[ht!]
\plotone{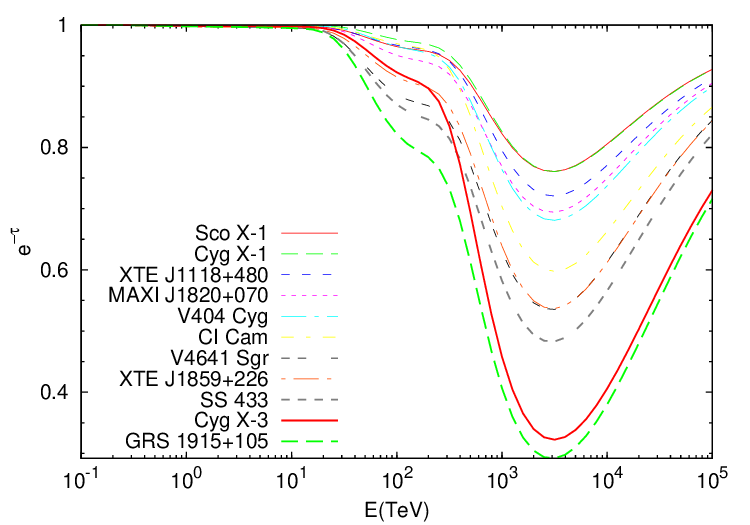}
\caption{Attenuation of the VHE gamma rays from selected Galactic microquasars, which are potential PeVatrons.
\label{fig2}}
\end{figure}

There are more than 20 microquasars are found in the Galaxy \citep{2016A&A...587A..61C,2023AA..675A.199A,2023AA..677A.134N}, some of them are very distant. In this paper, we selected the 11 microquasars(see table \ref{tab:binaries}) in LHAASO field of view to look at the attenuation effect in PeV region. Figure \ref{fig2} show the attenuation of gamma rays from these sources as a function of gamma ray energy. The attenuation of these sources are significant, especially for the distant sources Cyg X-3 and GRS 1915+105, the attenuation can reach up to 70\%.

\begin{deluxetable*}{lcccccccccccccc}
\tabletypesize{\scriptsize}
\tablewidth{0pt} 
\tablecaption{Selected Galactic microquasars as potential PeVatrons \label{tab:binaries}}
\tablehead{
\colhead{Source} & \colhead{Type} & \colhead{Dist(kpc)} &
\multicolumn{2}{c}{Coordinate} \\
\cline{4-5}
\colhead{} &
\colhead{} & \colhead{} & \colhead{$l$} & \colhead{$b$}
} 
\startdata 
{Sco X-1} & LMXB & 2.12 & 359.09 & 23.78 \\
 \hline
{Cyg X-1} & HMXB & 2.15 & 71.33 & 3.07 \\
 \hline
{XTE J1118+480} & LMXB & 2.57 & 157.66 & 62.32 \\
 \hline
{MAXI J1820+070} & LMXB & 2.81 & 35.85 & 10.16 \\
 \hline
{V404 Cyg} & LMXB & 3.01 & 73.12 & -2.09 \\
 \hline
{XTE J0421+560} & HMXB & 4.09 & 149.18 & 4.13 \\
 \hline
{V4641 Sgr} & LMXB & 4.74 & 6.77 & -4.79 \\
 \hline
{XTE J1859+226} & LMXB & 4.80 & 54.05 & 8.61 \\
 \hline
{SS 433} & LMXB & 5.50 & 39.69 & -2.24 \\
 \hline
{Cyg X-3} & HMXB & 8.95 & 79.85 & 0.70 \\
 \hline
{GRS 1915+105} & LMXB & 9.40 & 45.37 & -0.22 \\
\enddata
\tablecomments{We take the sources characteristics from papers(HMXB: \citep{2023AA..677A.134N} and LMXB \citep{2023AA..675A.199A}).}
\end{deluxetable*}

\subsection{The correction of PeVatron spectrum\label{subsec:figures}}
\begin{figure}[ht!]
\plotone{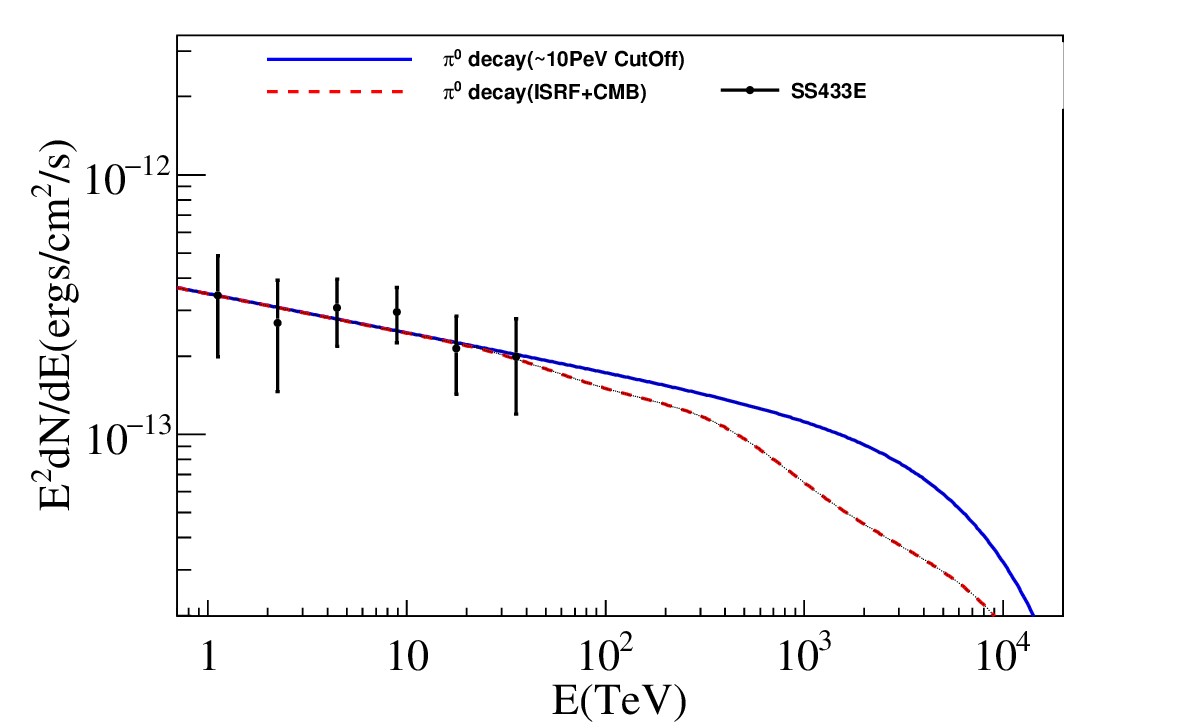}
\caption{Spectrum of the gamma ray emissions from SS 433E measured by the HESS telescope together with
calculated spectrum after the attenuation.
\label{fig3}}
\end{figure}

Based on the calculation above, the correction for the pair absorption can produces harder instrinsic spectra higher than 10 TeV than observed. To interpretate the origin of the gamma ray emission correctly from a source, this effect need to account. To see this, we selected several typical sources, the SS433 with distance 5.5 kpc, which has observed by HESS\citep{2024Sci...383..402H} and HAWC\citep{2018Natur.562...82A} above 10 TeV, and 1LHAASO J1959+1129u with distance 9.4kpc, which flux has published by LHAASO. The spectrums attributed to the PeVatron
should be ISRF-corrected. Figure \ref{fig3} shows the measured and ISRF-corrected data. The ISRF-corrected much higher than expected measurement, we can expect the LHAASO measurement in this energies in future.  Note that the hardening of the intrinsic spectrum around PeV may implicate different mechanism for the emissions.

LHAASO has detected dozens of PeVatrons. With the accumulation of LHAASO data, more distance PeVatrons are being measured. The correction will be important for assessing intrinsic spectral characteristics with emissions from these sources.

\begin{figure}[ht!]
\plotone{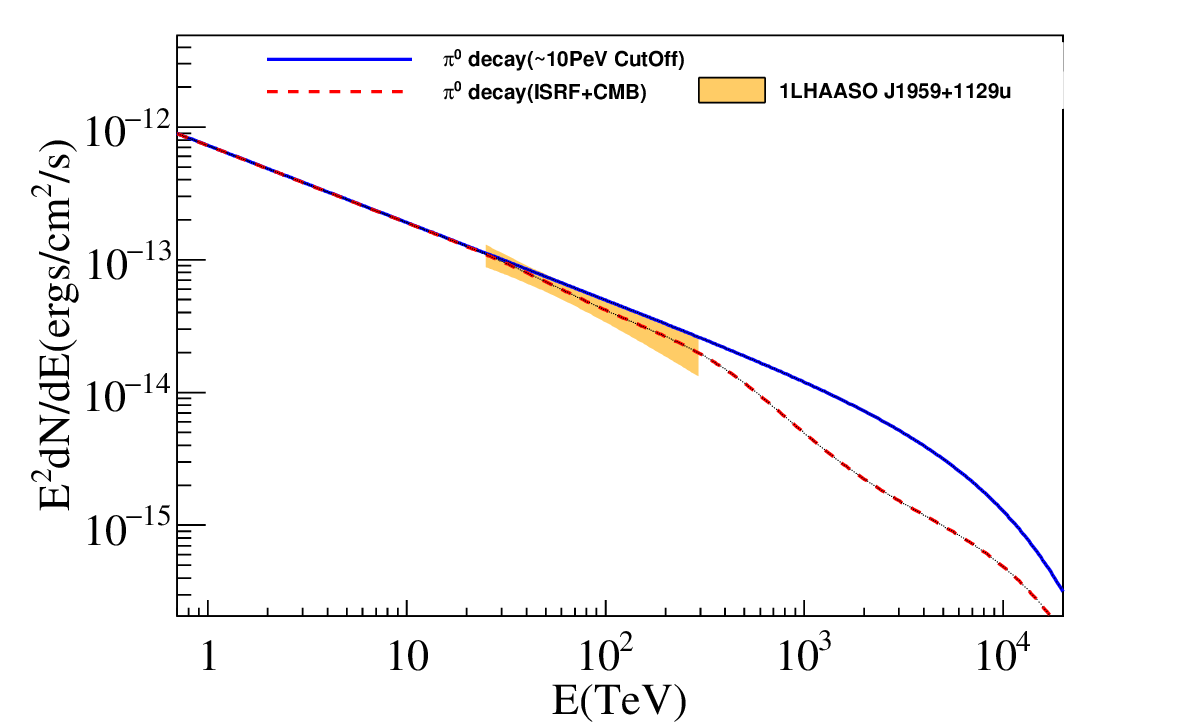}
\caption{Spectrum of the gamma ray emissions from  1LHAASO J1959+1129u measured by the LHAASO together with
calculated spectrum after the attenuation.
\label{fig4}}
\end{figure}

\section{Discussion} \label{sec:cite}
We used the ISRF model extracted from galprop code\citep{2022ApJS..262...30P} to calculate the absorption of gamma rays from PeV sources in first LHAASO catalogue by ISRF and CMB. Figure \ref{fig1} shows the relationship between the absorption effects of LHAASO sources and gamma energy, which is consistent with expectations. For more distant sources, such as 1LHAASO J1959+1129u with a distance of 9.4 kpc, the attenuation can reach up to 20\% at 100 TeV, and it can be as high as 70\% at about 3 PeV. This is very important to study the characteristics of the PeVatrons. The result may be affected by uncertainties in the source's distance, which could influence the correction magnitude. Additionally, the source's association may also have uncertainties. Nevertheless, this result is important, and studies of the LHAASO source catalog need to consider absorption to better understand the properties of the sources themselves.

We also calculate the attenuation effect of the microquasars in LHAASO field of view, which will be very important targets for LHAASO in future, to research the ultra high energy CRs acceleration capability of these sources. These calculation can be used to investigate the feature of emission model.

\begin{acknowledgments}
We would like to thank Dr. Hui Zhu for fruitful discussions, and Dr. LiJun Gou for carefully reading of the manuscript. This work is supported by Natural Sciences Foundation of China (No. 12375108 and 12275279).
\end{acknowledgments}

\bibliography{sample631}{}
\bibliographystyle{aasjournal}



\end{document}